\documentclass[aps,pra,preprint,groupedaddress,showpacs,superscriptaddress]{revtex4-1}
\usepackage{amsmath}
\usepackage[tight,footnotesize]{subfigure}
\usepackage[pdftex]{graphicx}
\newcommand{\mat}[1]{\mbox{\boldmath{$#1$}}} %To put greek letters in bold

\newcommand{\rmd}{{\mathrm{d}}}
\newcommand{\rme}{{\mathrm{e}}}
\newcommand{\rmi}{{\mathrm{i}}}

\begin{document}

%\preprint{Testando}

\title{Collective modes in free plasmas \\ subjected to a radiation field}


\author{B. V. Ribeiro}
%\email{brunovr@fis.unb.br}
\author{D. D. A. Santos}
%\thanks{}
\affiliation{Instituto de F\'{i}sica, Universidade de Bras\'{i}lia, CP: 04455, 70919-970 - Bras\'{i}lia - DF, Brasil}
\author{M. A. Amato}
%\email{maamato@unb.br}
\affiliation{Instituto de F\'{i}sica, Universidade de Bras\'{i}lia, CP: 04455, 70919-970 - Bras\'{i}lia - DF, Brasil}
\affiliation{International Center for Condensed Matter Physics,
Universidade de Bras\'{i}lia \\
CP: 04455, 70919-970 - Bras\'{i}lia - DF, Brasil}


\date{\today}

\begin{abstract}
In this study we report the effects of an external electromagnetic field on the collective properties of unmagnetized plasmas. The calculations are carried out in the semi-classical
approximation, \textit{i.e.}, the electromagnetic field is treated classically and the electrons from a
quantum mechanical viewpoint. The results show that the collective modes are damped away  smoothly and in a smaller frequency range than those reported by previous studies. An exponential-like decay for the plasmon frequencies as a function of the external field amplitude is readily observed. The results of previous studies are successfully obtained . We also find that the single photon processes has a pronounced effect on the decrease of the frequency range of modulation.

\end{abstract}

%\pacs{}
%\keywords{}

\maketitle

\section{Introduction}

Quantum-mechanical tools have been used extensively to deal with classical plasma physics phenomena for a long time \cite{PiSc62,WyPi62,harris}. One of the main reasons for the success of such an approach is that quantum perturbation theory provides a faster way to derive the equations describing classical plasmas, once the $\hbar\rightarrow 0$ limit is taken (see \cite{harris} and references therein for a concise summary of the motivations leading to the quantum-mechanical approach).

In recent years, special attention has been given to collective phenomena in quantum plasma systems \cite{BrMaMa2008,ShEl2010,VlTy2011}, in which high densities of the components and small average interparticle distances (of the order of the de Broglie wavelength) are assumed. In these works, a second quantization formalism is favored, leading to Wigner-Poisson and Wigner-Maxwell models. In references \cite{LaYu2011,LaYu2012}, the authors compute the dielectric response function and electronic conductivity of quantum and classical plasmas in the collisional regime using the Wigner-Boltzmann formalism \cite{VlTy2011}, while in reference \cite{LaYu2013}, the authors use the Schr\"odinger-Boltzmann formalism to write the dielectric function of a non-degenerate collisional plasma. In both formalisms, the collective response of the system is obtained through a perturbation on the chemical potential appearing in the usual Fermi-Dirac distribution for the equilibrium electrons, and good agreement is shown between the classical results and those of the quantum models with $\hbar\rightarrow 0$.

Another field of application for the quantum approach is that of laser-plasma systems \cite{adameck,amato,amato2}. In particular, photon-plasma interactions have been discussed by many authors (see \cite{harris} and references therein). In our work, we compute the dielectric response function of a collisionless, one-component plasma system with neutralizing background, interacting with an external radiation field. To do so, we avoid second quantization arguments and use the Schr\"odinger description to account for the state of the plasma constituents. We extend the work developed in \cite{adameck}, where there is no photon interaction contributing to the collective modes, in what they call a \textit{weak electron-radiation coupling}; and that developed in \cite{amato,amato2}, where two limiting cases are discussed, \textit{i.e.}, the strong-field limit, in which only multiphoton processes are significant, and the weak-field regime where only single-photon processes are significant.

For this study, the laser beam is treated as a classical plane electromagnetic wave in the dipole
approximation. We consider the laser linearly polarized along the $z$-axes, with the electric field
along the $x$-axes, taking into account a finite number
of photons interacting with the electron plasma. This simple extension gives rise to a different
dispersion relation for the electron waves. We see that an asymptotic value for the plasmon
frequency exists as we increase the radiation wave number, and that this frequency never goes
to zero, given a non-zero natural plasma frequency.
%The collective modes of the plasma are obtained numerically from the zeros of the dielectric function.

This paper is organized as follows: in section \rm{II}, we compute the state of the electrons through a unitary transformation \cite{galvao} using the Schr\"odinger formalism; in section \rm{III}, knowing the fluctuations in the wave function due to the external potential, we calculate the dielectric response function; in section \rm{IV}, we present the numerical scheme for obtaining the zeros of the dielectric function and discuss the results for the collective modes; in section \rm{V}, we calculate the electric conductivity; in section \rm{VI}, with an approximation on the plasmon frequency, we obtain an expression for the Landau damping term of the collective modes; we close with a summary of the main findings and difficulties.

\section{Electron States}

For an electron under the presence of an electromagnetic wave, the Schr\"odinger equation 
takes the form
\begin{equation}
	H \psi(\mathbf{r},t)=\rmi \hbar \frac{ \partial\psi(\mathbf{r},t)}{\partial t},
\label{sch}
\end{equation}
with the Hamiltonian operator
\begin{equation}
	H=\frac{1}{2m_e}(\mathbf{p} -e\mathbf{A}(t))^2 ,
\label{ham}
\end{equation}
where $\mathbf{p}$ is the momentum of the electron, $\mathbf{A}(t)$ is the vector potential of the radiation
\begin{equation}
	\mathbf{A}(t)=\left(\frac{E}{\omega}\sin(\omega t)\right)\hat{\mathbf{x}},
\label{vecpot}
\end{equation}
and $\omega$ is the frequency of the external radiation.
To solve this equation, given a time-dependent potential, we use a unitary transformation \cite{lima,galvao} of the form
\begin{equation}
	\psi\left(\mathbf{r},t\right) = U\Phi(\mathbf{r},t),
\label{trans}
\end{equation}
$\Phi(\mathbf{r},t)$ being the solution of the Schr\"odinger equation for a free electron and U the unitary operator given by
\begin{equation}
	U = \exp\left(\frac{\rmi}{\hbar}\mat{\alpha}(t)\cdot \mathbf{r}\right)exp\left(\frac{\rmi}{\hbar}\mat{\beta}(t)\cdot \mathbf{p}\right)\exp\left(\frac{\rmi}{\hbar}\eta(t)\right),
\label{u}
\end{equation}
where the functions $\mat{\alpha}(t), \mat{\beta}(t)$ produce, respectively, translations in momentum and space, and $\eta(t)$ is a phase factor. Substituting this into equation (\ref{sch}) we obtain
\begin{equation}
	\frac{\partial \psi(\mathbf{r},t)}{\partial t}=\frac{\rmi}{\hbar}\left(\frac{d\mat{\alpha}}{\rmd t}\cdot \mathbf{r} + \frac{\rmd\mat{\beta}}{\rmd t}\cdot \mathbf{p} + \frac{\rmd\eta}{\rmd t}\right) \psi(\mathbf{r},t) + U \frac{\partial \Phi(\mathbf{r},t)}{\partial t}.
\label{sch2}
\end{equation}

Multiplying this equation by $U^\dagger$, we get
\begin{eqnarray}
	H_0 &=& \frac{\mathbf{p}^2}{2m_e}+\frac{1}{2m_e}\left(\mat{\alpha}(t)-e\mathbf{A}(t)\right)^2 +\frac{\mathbf{p}}{m_e}\left(\mat{\alpha}(t)-e\mathbf{A}(t)\right) \nonumber \\
	&+& \frac{\rmd\mat{\alpha}}{\rmd t}\cdot \mathbf{r} +\frac{\rmd\mat{\beta}}{\rmd t}\cdot\mathbf{p}-\frac{\rmd\mat{\alpha}}{\rmd t} \cdot \mat{\beta} + \frac{\rmd\mat{\beta}}{\rmd t} \cdot \mat{\alpha} + \frac{\rmd\eta}{\rmd t},
\label{hzero}
\end{eqnarray}
where $H_0$ is the Hamiltonian for the free electron. We proceed to solve the equation for $\mat{\alpha}, \mat{\beta}$ and $\eta$ (linear terms in $\mathbf{r}$ and $\mathbf{p}$ and independent terms are set equal to zero) and find
\begin{equation}
	\psi(\mathbf{r},t) = \exp \left(\frac{\rmi}{\hbar} F(t)\right) \exp\left( \rmi \gamma_0 k_x \left( 1 - \cos(\omega t)\right) \right)
	\exp(\rmi\mathbf{k} \cdot \mathbf{r}) \exp \left( -\frac{\rmi}{\hbar} \varepsilon_\mathbf{k} t\right),
\label{psi}
\end{equation}
for the wave function of an electron in an electromagnetic field given by (\ref{vecpot}),
where we have simplified using $F(t) = -2\gamma_1 \omega t + \gamma_1 \sin(2\omega t)$, $\gamma_0 = \frac{eE}{m_e \omega^2}$, $\gamma_1 = \frac{e^2 E^2}{8m_e \omega^3}$ and $\varepsilon_\mathbf{k}$ is the energy of the free electron with wave number $\mathbf{k}$.

We can see that (\ref{psi}) forms an orthonormal set. For a plasma, the presence of the external field generates local fluctuations in   potential, and we may use (\ref{psi}) as basis to expand the wave function of these electrons in a local potential 
\begin{equation}
	\Psi_k(\mathbf{r},t) = \sum_{\mathbf{k}} a_{\mathbf{k}}(t) \psi(\mathbf{r},t).
\label{sum}
\end{equation} 

Assuming a weak local potential of the form
\begin{equation}
	\varphi(\mathbf{r},t)=\int \rmd\mathbf{q} \int \rmd\Omega \exp(\rmi \mathbf{q} \cdot \mathbf{r}) \exp( -\rmi \Omega t) \varphi(\mathbf{q},\Omega) + \mathrm{c.c.}
\label{local}
\end{equation}
we determine the coefficients $a_{\mathbf{k}}(t)$ using usual perturbation theory \cite{sakurai}
\begin{multline}
	a_{\mathbf{k+q}}(t) = e \cdot \exp(-\rmi \gamma_0 q_x) \sum_{\mathrm{m},\Omega}\rmi^\mathrm{m} \mathrm{J}_\mathrm{m} (q_x \gamma_0) \varphi(\mathbf{q},\Omega) \\
	\frac{\exp\left(\frac{\rmi}{\hbar} \left( \varepsilon_{\mathbf{k+q}} - \varepsilon_{\mathbf{k}} - \hbar \Omega-\mathrm{m}\hbar \omega \right) t \right)}{\varepsilon_{\mathbf{k+q}} - \varepsilon_{\mathbf{k}} - \hbar \Omega-\mathrm{m}\hbar \omega - \rmi\zeta} ~~ (\zeta \rightarrow 0^+), 
\label{aaa}
\end{multline}
where we use the identity
\begin{equation}
	\exp(\rmi\alpha \cos x) = \sum_{\mathrm{m}} \rmi^\mathrm{m} \mathrm{J}_\mathrm{m} (\alpha)\exp(\rmi \mathrm{m}x),
\label{bessel}
\end{equation}
with $\mathrm{J}_\mathrm{m} (\alpha)$ being the Bessel function of order $\mathrm{m}$.

Finally, we can write the wave function as
\begin{multline}
	\Psi_k(\mathbf{r},t)=\rme^{\frac{\rmi}{\hbar} F(t)} \exp\left( \rmi \gamma_0 k_x \left( 1 - \cos(\omega t) \right)\right)
	\\ 
	\rme^{-\frac{\rmi}{\hbar} \varepsilon_{\mathbf{k}}t + \rmi\mathbf{k} \cdot \mathbf{r}} \left\{ 1 + e \sum_{\mathbf{q} , \Omega} \exp \left( -\rmi \gamma_0 q_x \cos(\omega t)\right) \right. \\
	\left. \sum_\mathrm{m} \rmi^\mathrm{m} \mathrm{J}_\mathrm{m} (q_x \gamma_0) \frac{\varphi(\mathbf{q},\Omega) \rme^{ -\rmi (\Omega + \mathrm{m} \omega)t}}{\varepsilon_{\mathbf{k+q}} - \varepsilon_{\mathbf{k}} - \hbar\Omega - \mathrm{m} \hbar\omega - \rmi\zeta} \rme^{\rmi\mathbf{q} \cdot \mathbf{r}} \right\}
\label{psitot}
\end{multline}

Equation (\ref{psitot})  is the wave function for the electrons of the plasma under the incidence of the radiation given by (\ref{vecpot}).

\section{Dielectric Function}

Knowing the states of the electrons via (\ref{psitot}), we can obtain the fluctuation of the charge density
\begin{equation}
	\rho_k (\mathbf{r},t) = -e\left|\Psi_k(\mathbf{r},t)\right|^2 - \rho_k ^{(0)}(\mathbf{r},t),
\label{fluc}
\end{equation}
$\rho_k ^{(0)}(\mathbf{r},t)$ being the charge density in the absence of a weak local potential, \textit{i.e}, the charge density given by the $\psi(\mathbf{r},t)$ distribution. Neglecting terms of higher orders in $\varphi$, we have
\begin{multline}
	\rho_k (\mathbf{r},t)=-e^2 \sum_{\mathbf{q},\Omega} \rme^{ \rmi\mathbf{q} \cdot \mathbf{r}} \varphi(\mathbf{q},\Omega) \exp( -\rmi\gamma_0 q_x \cos(\omega t)) \\
	\sum_\mathrm{m} \rmi^\mathrm{m} \mathrm{J}_\mathrm{m} (q_x \gamma_0) \left\{ \frac{\rme^{ -\rmi(\Omega + \mathrm{m} \omega)t}}{\varepsilon_{\mathbf{k+q}} -\varepsilon_{\mathbf{k}} - \hbar\Omega - \mathrm{m} \hbar\omega - \rmi\zeta}\right. \\
	\left. + \frac{\rme^{ \rmi(\Omega + \mathrm{m} \omega)t}}{\varepsilon_{\mathbf{k+q}} -\varepsilon_{\mathbf{k}} + \hbar\Omega +\mathrm{m}\hbar\omega + \rmi\zeta}\right\}.
\label{flucsum}
\end{multline}

Assuming a Maxwellian distribution $f_{\mathbf{k}}$ for the electrons ( for a discussion on the choice of this distribution see ref \cite{GuMiFoAgNu2007}), we have the total fluctuation as
\begin{equation}
	\rho(\mathbf{r},t) = \sum_{\mathbf{k}} f_{\mathbf{k}} \rho_k (\mathbf{r},t)
\label{rho}.
\end{equation}
Using, once more, relation (\ref{bessel}), we obtain
\begin{equation}
	\rho(\mathbf{r},t) = -\rme ^2 \sum_{\mathbf{q},\Omega} \rme^{\rmi \mathbf{q} \cdot \mathbf{r} - \rmi\Omega t} \varphi(\mathbf{q},\Omega)
	\sum_{\mathrm{m},\mathrm{m}'} \rmi ^{\mathrm{m} - \mathrm{m}'} \mathrm{J}_{\mathrm{m}} (q_x \gamma_0) \mathrm{J}_{\mathrm{m'}} (q_x \gamma_0)\rme^{-\rmi(\mathrm{m} - \mathrm{m}') \omega t} \Pi(\mathbf{q},\Omega + \mathrm{m} \omega),
\label{rhotot}
\end{equation}
where $\Pi(\mathbf{q},\Omega)= \sum_{\mathbf{k}} \frac{f_{\mathbf{k+q}} - f_{\mathbf{k}}}{\varepsilon_{\mathbf{k+q}} - \varepsilon_{\mathbf{k}} -\hbar \Omega - \rmi\zeta}$ is the electronic polarizability and $\mathrm{m}$ corresponds to the number of photons involved in the process \footnote{Notice how $\mathrm{m}$ appears as the number of $\hbar \omega$ contributing to the polarizability}. We take the real part of (\ref{rhotot}) to calculate the fluctuation. This fluctuation induces a potential in the medium given by the Poisson equation
\begin{equation}
	\nabla^2 \varphi_{\mathrm{ind}} (\mathbf{r},t) = -4\pi \rho(\mathbf{r},t).
\label{poison}
\end{equation}

Using (\ref{rhotot}) and the Fourier transform of (\ref{poison}), we obtain
\begin{equation}
	\varphi_{\mathrm{ind}} (\mathbf{q},\Omega) = \frac{4\pi e^2}{q^2}\varphi(\mathbf{q},\Omega) \sum_\mathrm{m} \mathrm{J}_\mathrm{m} ^2 (q_x \gamma_0) \Pi(\mathbf{q},\Omega + \mathrm{m}\omega),
\label{potind}
\end{equation}
which is the induced part of the \textit{full} local potential
\begin{equation}
	\varphi(\mathbf{q},\Omega) = \varphi_{\mathrm{ext}} (\mathbf{q},\Omega) + \varphi_{\mathrm{ind}} (\mathbf{q},\Omega) = \frac{\varphi_{\mathrm{ext}} (\mathbf{q},\Omega)}{\epsilon(\mathbf{q},\Omega)},
\label{fullpot}
\end{equation}
where $\epsilon(\mathbf{q},\Omega)$ and $ \varphi_{\mathrm{ext}} (\mathbf{q},\Omega)$ are the dielectric function of the plasma and the external potential, respectively. The roots of the real part of this dielectric function give us the frequency of the longitudinal waves (collective oscillations, \textit{i.e}, plasmons) in the plasma. We, therefore, separate $\epsilon(\mathbf{q},\Omega)$ into a real ($\epsilon_{\mathrm{R}}$) and an imaginary ($\epsilon_{\mathrm{I}}$) part. We proceed to the classical limit by letting $\hbar\rightarrow 0$ and, after some algebraic work, we are left with
\begin{equation}
	\epsilon_{\mathrm{R}}(\mathbf{q},\Omega) = 1 - \omega_p ^2 \sum_\mathrm{m} \mathrm{J}_\mathrm{m} ^2 (q_x \gamma_0)\frac{1}{\lambda^2}
	\cdot\left(1 + \frac{2\left\langle \mathbf{q . v} \right\rangle}{\lambda} +\frac{ 3\left\langle (\mathbf{q . v})^2 \right\rangle}{\lambda^2}\right)\rme^{\left( -\frac{\varepsilon_{\gamma}}{k_B T}\right)},
\label{realpart}
\end{equation}
where $\lambda=\mathrm{m}\omega + \Omega$, $\omega_p$ is the plasma natural frequency, $\varepsilon_{\gamma} = 2\gamma_1 \omega$ is the energy of the electromagnetic radiation and the average $\left\langle g(\mathbf{q,v})\right\rangle$ is taken with respect to the function $f_k$.

\section{Collective Modes}

\begin{figure}[h]

\center
\subfigure{\includegraphics[width=8cm]{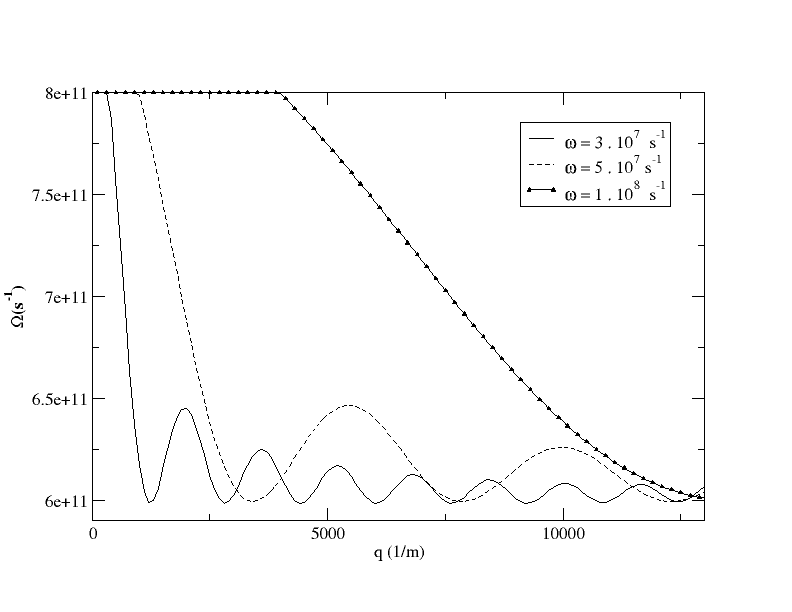}}
\subfigure{\includegraphics[width=8cm]{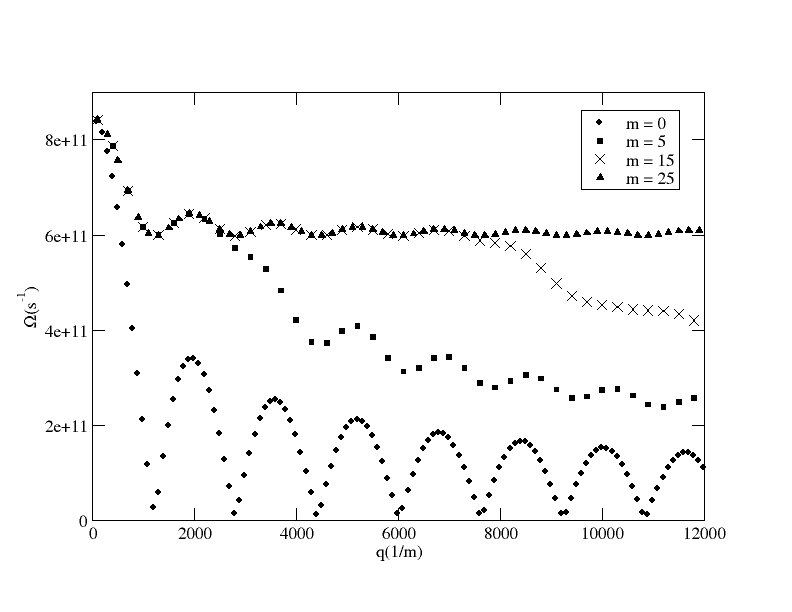}}
\caption{\label{disp}(left) The natural plasma frequency for this case is $6.10^{11} s^{-1}$ and $E=10 V/m$. Notice that, as we increase the radiation frequency, the dispersion decays slower. (right) There is a assymptotic value for the plasmon frequency as we increase both $\mathrm{m}$ an $q$, which seems to coincide with the natural frequency. The curve for $\mathrm{m}=0$ reproduces exactly that of ref. \cite{adameck}.}

\end{figure}

\begin{figure}[h]

\center
\subfigure{\includegraphics[width=8cm]{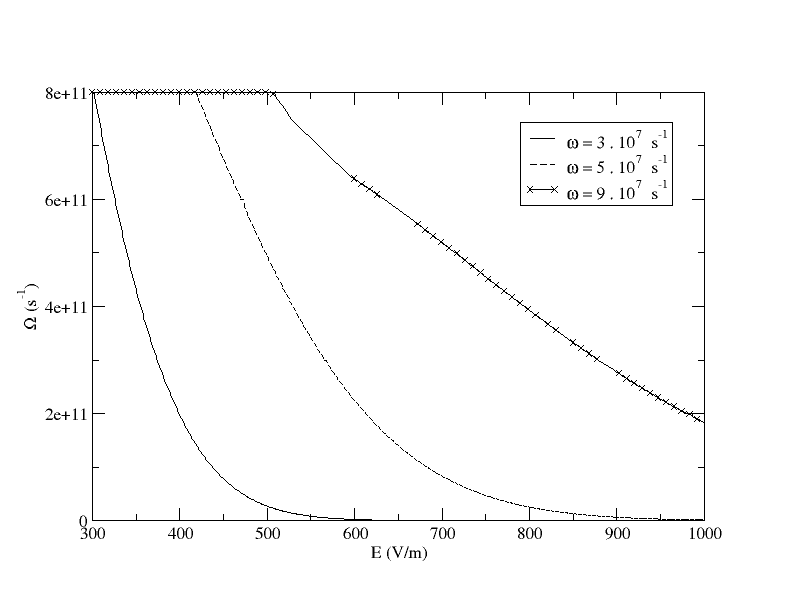}}
%\qquad
\subfigure{\includegraphics[width=8cm]{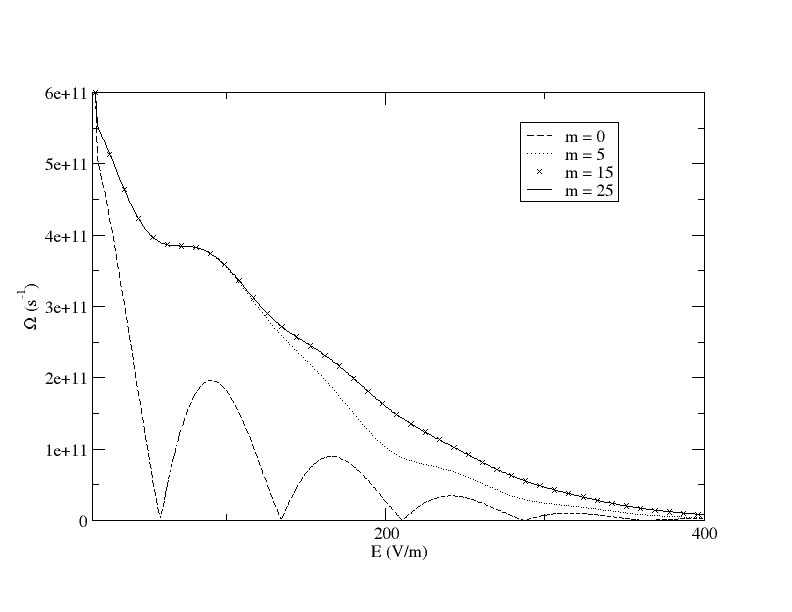}}
\caption{\label{ecamp}(left) The natural plasma frequency for this case is, also, $6.10^{11} s^{-1}$ and the value of $q$ is generated randomly in the interval of figure(\ref{disp}).For the $E$ dependence, we see an almost exponential-like decay, which agrees with the exponential term that appears in (\ref{realpart}). (right) We see the tendency of the curve to become an exponential-like decay as we increase the number of photons in the process.}

\end{figure}

The roots of (\ref{realpart}) give us the frequencies of collective oscillations in the plasma as a function of their wave number ($\Omega(\mathbf{q})$). Here we encounter a numerical difficulty, as the sumation over $\mathrm{m}$ does not allow us to get an analytical result for this frequencies. However, if we fix a value for $\mathbf{q}$, we obtain an expression depending only on $\Omega$, which is easily solvable using a bisection method in FORTRAN language \cite{nr}. The sum over $\mathrm{m}$ is truncated for a value of $\mathrm{m}$ beyond which the terms become neglegible. In this way, we take values of q from zero to about $20000\mathrm{m}^{-1}$, and for each $q$ we get a frequency value. So, we can plot the dispersion relation ($\Omega$ versus $q$) of this system for various values of the radiation frequency. In the same manner, we can fix values of $E$ and get the dependece of $\Omega$ with $E$. These plots are shown in figures (\ref{disp}) and (\ref{ecamp}). We see that, for any given value of $\omega$ (the frequency of the radiation), the asymptotic value for the plasmon frequency is the same, the difference is that for larger values of $\omega$ the plasmon frequency decays slower. From this we see that, for a radiation with large enough energy, the plasma remains unperturbed.

We can also vary the number of photons involved in the process and see how the curves react for a natural plasma frequency of $6.10^{11}s^{-1}$. This plots are shown in figures (\ref{disp}) and (\ref{ecamp}).

In figure (\ref{disp}) wee note a clear restriction to the range of frequencies allowed to the plasmons as $\mathrm{m}$ increases. So, as more photons participate in the process, more energetic the plasmons are (given the same value of $q$). For all plots, the temperature of the plasma is $k_B T = 1,6 . 10^{-19} J$.

The dispersion relation obtained for the free plasma subjected to a radiation field is
\begin{equation}
	\epsilon_\mathrm{R}(\mathbf{q},\Omega) = 0
\label{disprealfin}
\end{equation}
\vspace{0.2cm}
\begin{equation}
	1 - \omega_p ^2 \sum_\mathrm{m} \mathrm{J}_\mathrm{m} ^2 (q_x \gamma_0) \frac{1}{\lambda_\mathrm{m}^2} \left( 1 + \frac{3q^2\left\langle v^2 \right\rangle}{\lambda_\mathrm{m}^2} \right) \rme^{ -\frac{\varepsilon_{\gamma}}{k_B T_e}} = 0.
\label{disprelfinal}
\end{equation}

This equation is quadratic in $\Omega$. We can easily check (\ref{disprelfinal}) to confirm that there are pairs of solutions of the form $\pm \Omega$. Furthermore, this solutions must also be even functions of $q$.

\begin{figure}[h]

\center

\includegraphics[width=9cm]{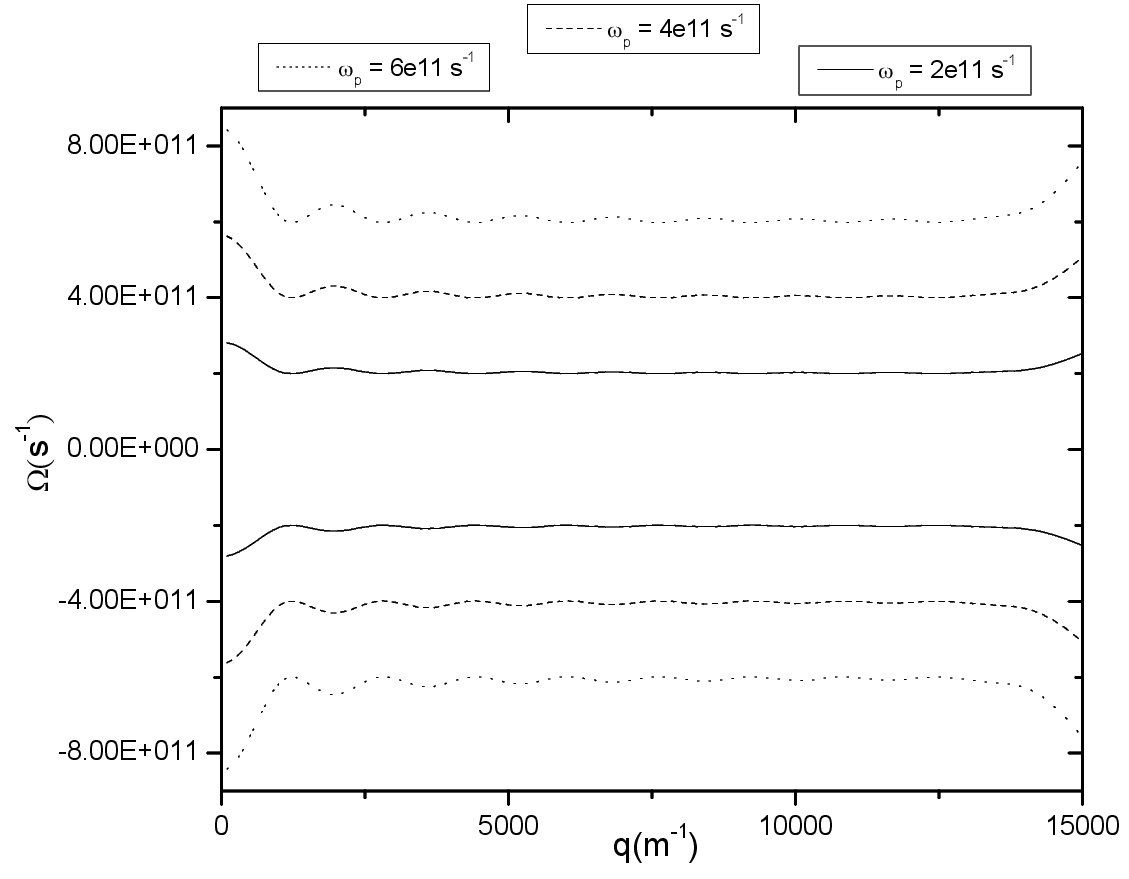}
\caption{\label{disp2}Plot obtained with $E = 10 V/m$ and $\omega = 3 \cdot 10^7 s^{-1}$. Dispersion relation for various values of the plasma frequency. The modes excited in plasmas with lower natural frequencies have lower energies.}

\end{figure}

Using typical parameters of discharge plasma we plot in figure (\ref{disp2}) the dispersion relation for a radiation field in the radiofrequency range.

%\begin{figure}[h]
%
%\center
%\subfigure{\includegraphics[width=8cm]{novas/disprelrange.pdf}}
%\qquad
%\subfigure{\includegraphics[width=8cm]{BW/negepos4.png}}
%\caption{\label{disp2}Plot obtained with $E = 10 V/m$ and $\omega = 3 \cdot 10^7 s^{-1}$.(top) Dispersion relation for a discharge plasma subjected to radiofrequency radiation. (bottom) Dispersion relation for various values of the plasma frequency. The modes excited in plasmas with lower natural frequencies have lower energies.}
%
%\end{figure}

As expected, for lower plasma frequencies, the plasmon frequency assumes smaller values as well, retaining the assymptotic value at the plasma frequency. We must note that for large $\left|q\right|$, the code presents strong numerical fluctuations. Therefore the behavior of the curve in this region can not be attributed to any physical cause.

\begin{figure}[h]

\center
\subfigure{\includegraphics[width=5.5cm]{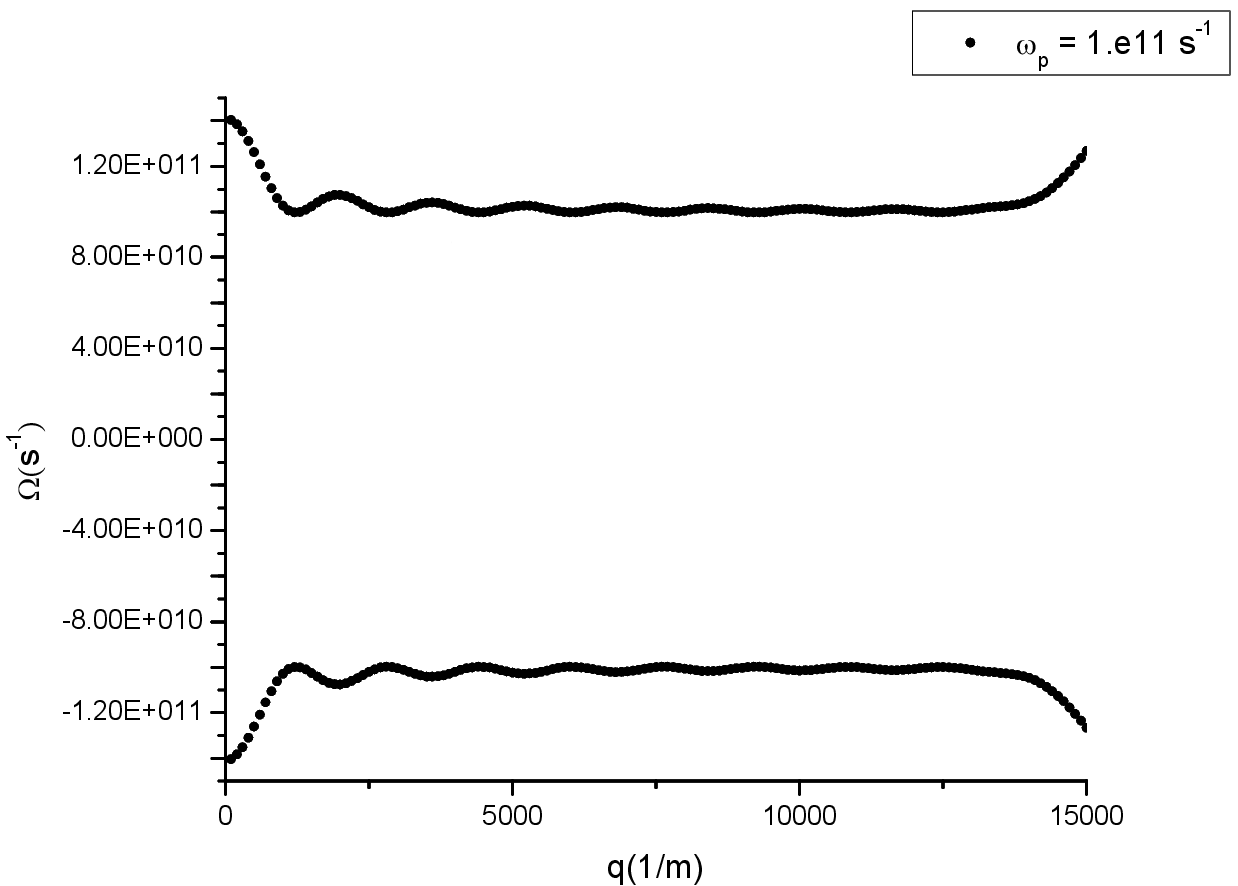}}
\quad
\subfigure{\includegraphics[width=5.5cm]{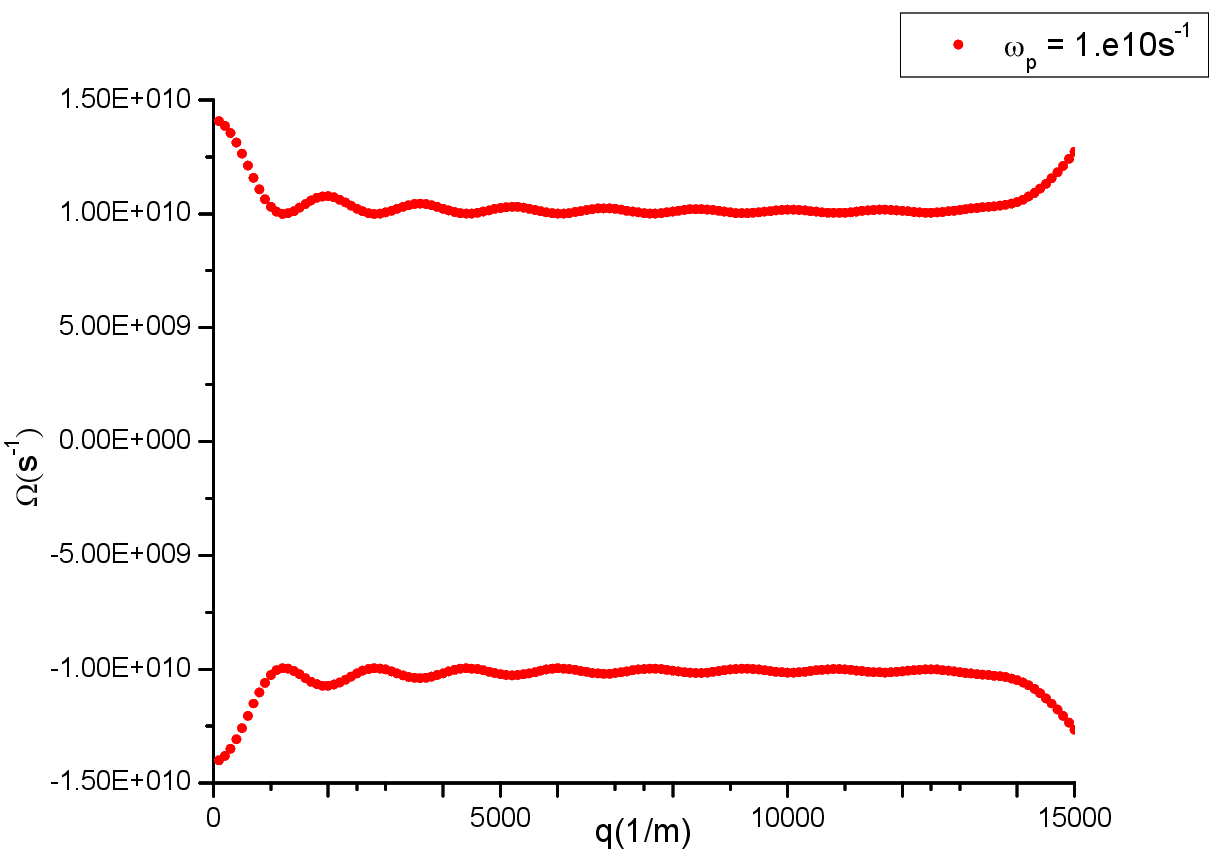}}
\subfigure{\includegraphics[width=5.5cm]{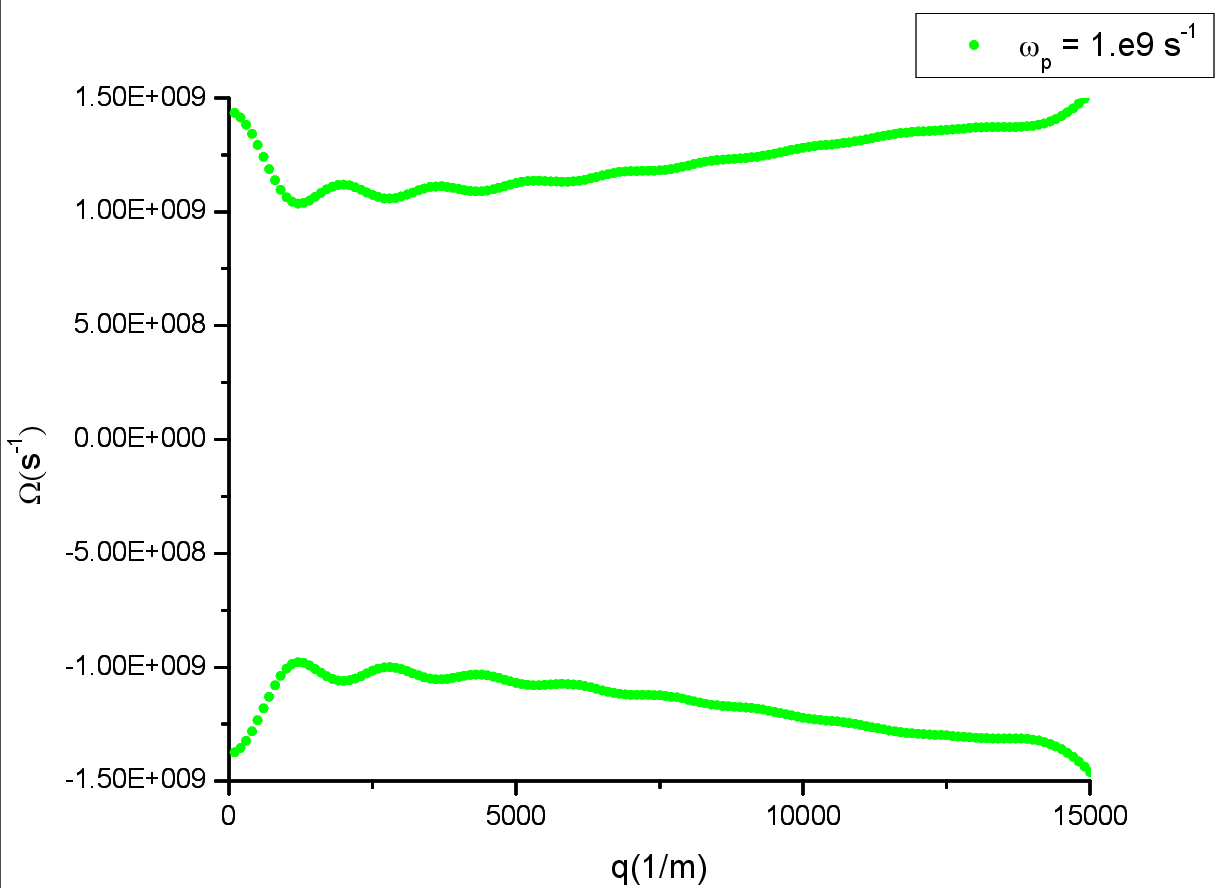}}
\caption{\label{1e10}(left) Dispersion relation for $\omega_p=1.e11 s^{-1}$. (right) Dispersion relation for $\omega_p=1.e10 s^{-1}$. (bottom) Dispersion relation for $\omega_p=1.e19 s^{-1}$.}

\end{figure}

If we assume smaller values for the plasma frequency, we can observe numerical instabilities for much smaller values of the plasmon wave number, as shown in the graphs of figure (\ref{1e10}).

\begin{figure}[h]
	\centering
		\includegraphics[width=8cm]{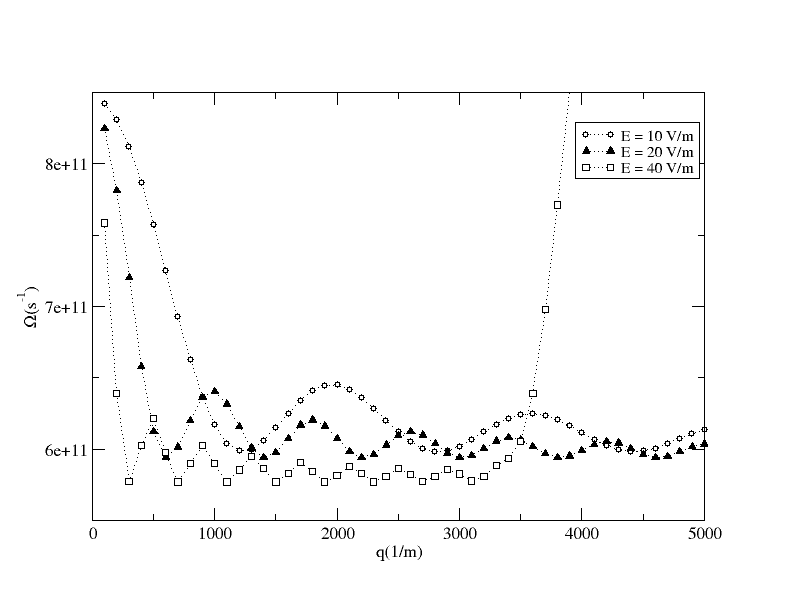}
	\caption{Dispersion relation for various external field amplitudes.}
	\label{variandoE}
\end{figure}

Taking only the first quadrant, we examine the dependence of the dispersion relation with the amplitude of the external field (figure (\ref{variandoE}) ). For large values of $E$, the curve gets \textit{rougher} and the numerical instabilities happen at smaller values of $q$. Actually, it is not yet clear if this instabilities are due to numerical fluctuations or to a break down of our linear assumption.

\begin{figure}[h]

\center
\subfigure{\includegraphics[width=8cm]{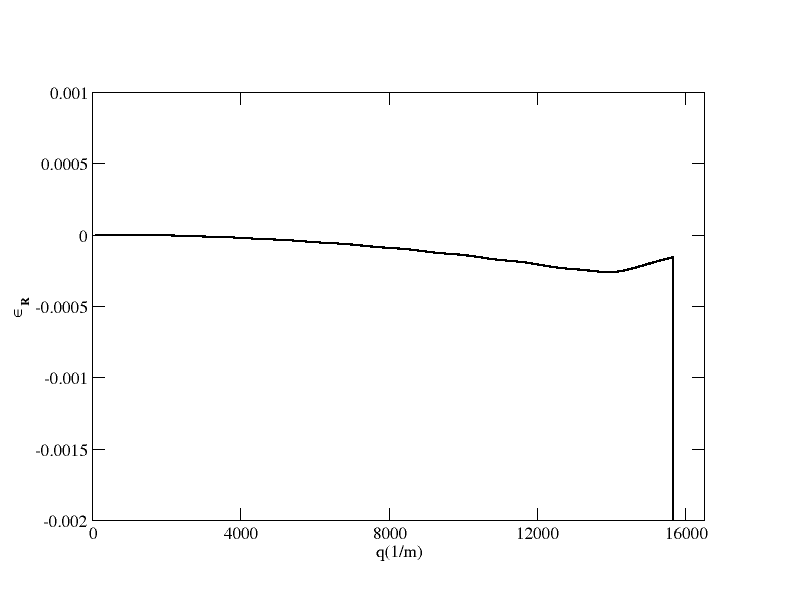}}
\subfigure{\includegraphics[width=8cm]{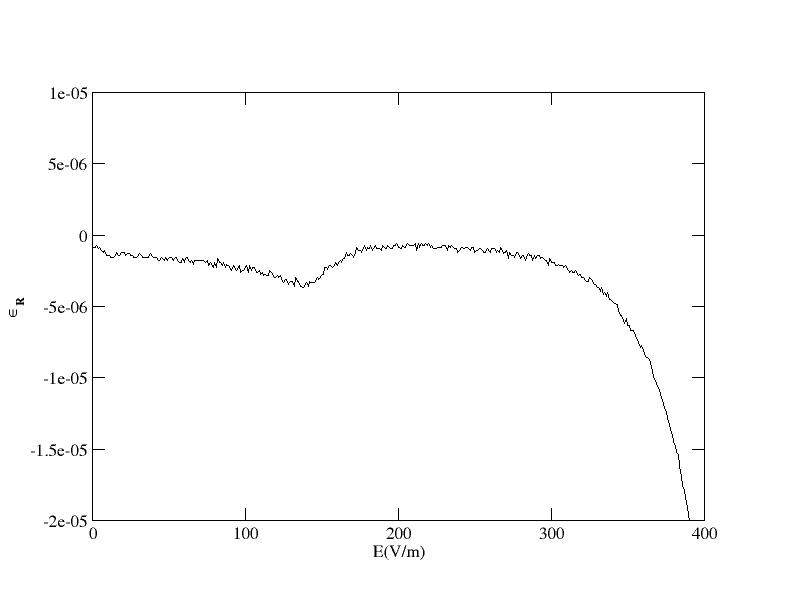}}
\caption{\label{qtest}Real part of the dieletric function for the collective modes: (left) Fixed $E=10V/m$, (right) Fixed $q=210m^{-1}$.}

\end{figure}

To check the validity of our code, we can easily plot the value of $\epsilon_R(q,\Omega)$ to see if it is actually close to zero. We observe in figure (\ref{qtest}) that our code shows consistent results. For larger values of $\left|q\right|$ and $|E|$, the results get further away from the actual roots of $\epsilon_R(q,\Omega)$.

We can, also, estimate the imaginary part of the dieletric function in the classical limit.
\begin{equation}
	\epsilon_\mathrm{I}(\mathbf{q},\Omega) = \sqrt{\frac{\pi}{2}}\left( \frac{m_e}{k_B T} \right)^{3/2} \frac{\omega^2_p}{q^3} \rme^{-\varepsilon_{\gamma}/k_B T}
	\sum_\mathrm{m} \mathrm{J}^2_\mathrm{m}(q\gamma_0) \lambda_\mathrm{m} \rme^{ -\frac{m_e (\lambda_\mathrm{m}/q)^2}{2k_B T}}
\label{final}
\end{equation}

This expression will be usefull in calculating the dynamical conductivity, in the spirit of \cite{Adetal2009}, in the next section.

%For given plasma frequency and external electric field amplitude, we plot values of $\epsilon_\mathrm{I}(\mathbf{q},\Omega)$ as function of $q$.
%
%\begin{figure}[h]
%	\centering
%		\includegraphics[width=8cm]{novas/dielImag.pdf}
%	\caption{Imaginary part of the dielectric function.}
%	\label{dielImag}
%\end{figure}
%
%We can observe, cearly, a peak of the function around the region of $q$ corresponding to propagative colletive behavior (excited plasmons).

\section{Conductivity}

The conductivity of the system can be obtained from the expression
\begin{equation}
	\epsilon(q,\Omega) = 1 + \frac{4\pi \rmi}{\Omega} \sigma(q,\Omega)
\label{cond1}
\end{equation}

If we write the conductivity as $\sigma=\sigma_\mathrm{R} + \rmi \sigma_\mathrm{I}$, we arrive at the relations
\begin{eqnarray}
	\sigma_\mathrm{R}(q,\Omega) &=& \frac{\Omega}{4\pi} \epsilon_\mathrm{I}(q,\Omega) \nonumber \\
 \\
	\sigma_\mathrm{I}(q,\Omega) &=& \frac{\Omega}{4\pi}\left[1 - \epsilon_\mathrm{R}(q,\Omega)\right] \nonumber
\label{condri}
\end{eqnarray}

\begin{figure}[h]
	\centering
		\includegraphics[width=9cm]{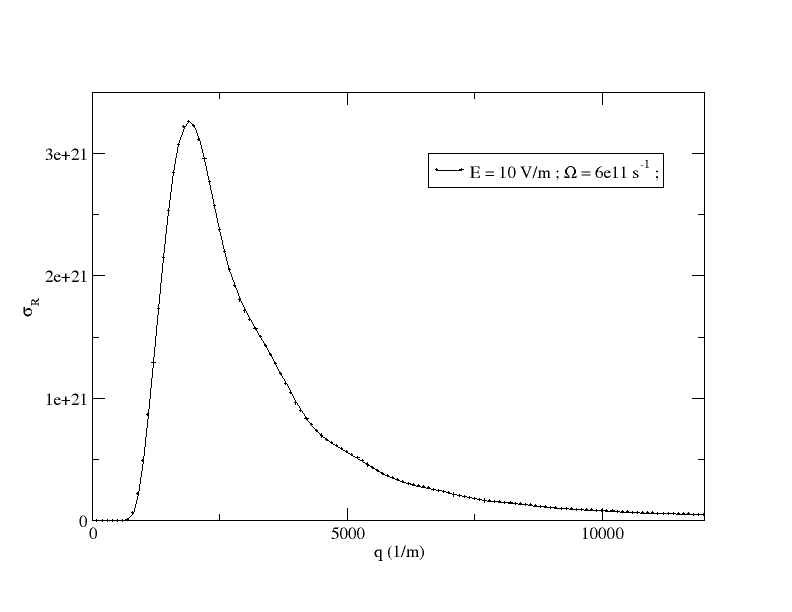}
	\caption{Real part of the conductivity for various values of $q$.}
	\label{sigreal}
\end{figure}

The real part of the conductivity presents a strong $q$ dependent peak, shown in figure \ref{sigreal}

\begin{figure}[h]
	\centering
		\includegraphics[width=9cm]{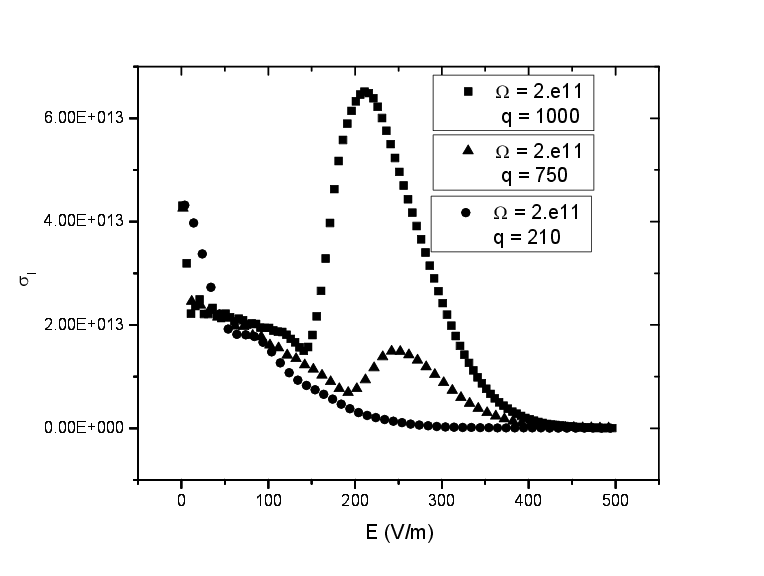}
	\caption{Imaginary part of the conductivity for various values of $q$.}
	\label{sigme}
\end{figure}

For a fixed $\Omega$, we can see, in figure (\ref{sigme}) an oscillatory behavior in the imaginary part of the conductivity for small values of $E$. A strong, $q$ dependent, peak is followed by a damped region. We see that $\sigma_\mathrm{I}$ goes to zero in the region of collective behavior. (See figure (\ref{ecamp})).

\begin{figure}[h]
	\centering
		\includegraphics[width=9cm]{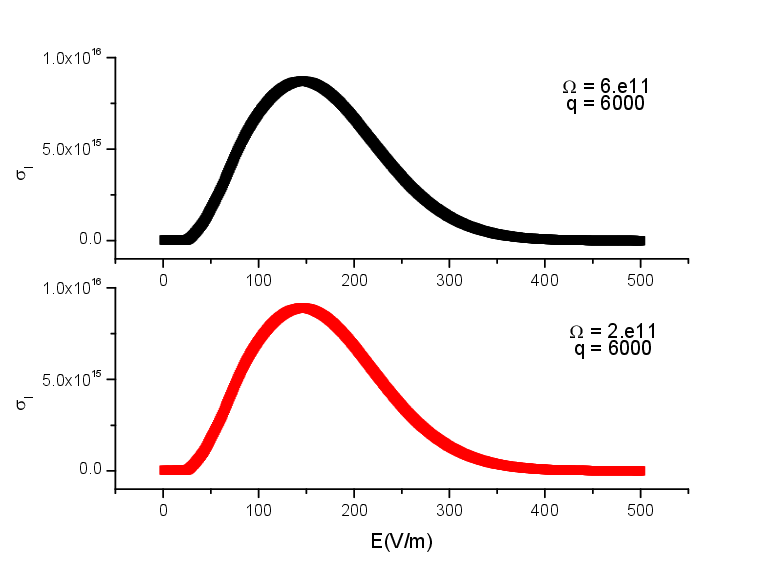}
	\caption{$\sigma_\mathrm{I}$ for slightly diferent values of $\Omega$.}
	\label{sigidifo}
\end{figure}

As expected from eq. (\ref{condri}), a small change \footnote{We use the term \textit{small change} to refer to a change in the value that does not change the order of magnitude.} in $\Omega$ does not change significantly the value of $\sigma_\mathrm{I}$ as can be seen from figure (\ref{sigidifo}).

\section{Landau term}

So far we have assume a real value for the plasmon frequency $\Omega$. Nevertheless, the dielectric function was defined in a such a way that it can assume complex values ($\epsilon= \epsilon_\mathrm{R} + \rmi \epsilon_\mathrm{I}$) for real $q$ and $\Omega$. \textit{For complex  $\Omega$, $\epsilon(q,\Omega)$ is to be interpreted as the analytic continuation of $\epsilon(q,\Omega)$ from the real $\Omega$ axis.} \cite{harris}.

We write the solution of the equation $\epsilon=0$ in the form
\begin{equation}
	\Omega=\Omega_\mathrm{R} + \rmi\gamma.
\label{landter}
\end{equation}

If we assume 
\begin{equation}
	\left| \gamma \right| << \left| \Omega_\mathrm{R} \right|, \quad \left|\epsilon_\mathrm{I}\right| << \left|\epsilon_\mathrm{R}\right| .
\label{gamma}
\end{equation}

we can expand $\epsilon=0$ and neglect terms of higher order in small terms to obtain
\begin{equation}
	\epsilon_\mathrm{R} (q,\Omega_\mathrm{R}) + \rmi\gamma \left(\frac{\partial \epsilon_\mathrm{R}}{\partial \Omega}\right)_{\Omega = \Omega_\mathrm{R}} + \rmi\epsilon_\mathrm{I}(q,\Omega_\mathrm{R}) = 0.
\label{expterm}
\end{equation}

Thus, we get two equations for the real and imaginary parts
\begin{eqnarray}
	\epsilon(q,\Omega_\mathrm{R}) &=& 0 ~~~~\leftarrow \mathrm{Used~in~previous~sections} \nonumber \\
\nonumber \\
	\gamma &=& -\frac{\epsilon_\mathrm{I}(q,\Omega_\mathrm{R})}{\left( \frac{\partial\epsilon_\mathrm{R}}{\partial\Omega} \right)_{\Omega_\mathrm{R}}}
\label{2parts}
\end{eqnarray}

Using our results, and neglecting higher orders in $(1/\lambda_\mathrm{m})$, we obtain
\begin{equation}
	\gamma = - \left( \frac{m_e}{2k_B T} \right)^{3/2}\frac{\sqrt{\pi}}{q^3}\frac{ \sum_\mathrm{m} \mathrm{J}^2_\mathrm{m}(q\gamma_0) \lambda_\mathrm{m} \exp\left( -\frac{m_e(\lambda_\mathrm{m}/q)^2}{2k_B T} \right) }{\sum_\mathrm{m} \mathrm{J}^2_\mathrm{m}(q \gamma_0) \lambda^{-3}_\mathrm{m}}
\label{landau}
\end{equation}
for the imaginary part of the plasmon frequency, responsable for the damping (or growth) of the collective modes.

\section{Conclusion}
The number os photons involved in the interaction between the plasma and the electromagnetic radiation is of extreme importance. We see that a single photon restricts the range of frequencies allowed to the plasmons (see figure (\ref{disp})), and, as we increase the number of photons, an assymptotic value appears for these frequencies.

The plasmon frequency decays very rapidly with increasing field amplitude ($E$) in an exponential-like curve (see figure(\ref{ecamp})). As we increase the number of photons in the process, we no longer see an oscillation in the curve of $\Omega$ versus $E$ as reported by \cite{adameck}.

We successfully derived an equation for the conductivity of the system and a first approximation for the imaginary part of the plasmon frequency.

Although the codes present strong numerical fluctuations in regions of high valued electric field amplitude, the divergences of the dispersion relation in these regions can, also, be associated with the break down of the linear approximation (as well as the non-relativistic one). The effects of nonlinearty are under study and we are currently extending these results to compute magnetic field effects.

\begin{acknowledgments}
The authors D.D.A.S. and B.V.R. are grateful to CAPES for grant of scholarship during the course of this work. M.A.A. thanks CNPq for the financial support.
\end{acknowledgments}

\bibliography{apssamp}

\end{document}